\documentclass[a4paper,onecolumn,12pt,assc]{IEEEtranASSC}
\usepackage{graphicx}
\usepackage{subcaption}

\usepackage{float}
\restylefloat{figure}

\usepackage{url}

\begin{document}
\newcommand{\bruu}{{\sc{bats-r-us}}}
\newcommand{\bru}{{\sc{bats-r-us }}}	
\newcommand{\tb}{{$\tau$ Boo }}
\newcommand{\tbb}{{$\tau$ Boo}}
	
%
\title{Winds of Planet Hosting Stars}

\include{ASRC14_StarWinds.tex}

\author{\IEEEauthorblockN{B. A. Nicholson\IEEEauthorrefmark{1},
M. W. Mengel\IEEEauthorrefmark{1},
L. Brookshaw\IEEEauthorrefmark{1},
A. A. Vidotto\IEEEauthorrefmark{2},
B. D. Carter\IEEEauthorrefmark{1},
S. C. Marsden\IEEEauthorrefmark{1},
J. Soutter\IEEEauthorrefmark{1},
I. A. Waite\IEEEauthorrefmark{1} and
J. Horner\IEEEauthorrefmark{1}\IEEEauthorrefmark{3}}

\IEEEauthorblockA{\IEEEauthorrefmark{1}
Computational Engineering and Science Research Centre, University of Southern Queensland, Toowoomba, QLD 4350, Australia}

\IEEEauthorblockA{\IEEEauthorrefmark{2}
Geneva Observatory, University of Geneva, Sauverny, Versoix CH-1290, Switzerland}

\IEEEauthorblockA{\IEEEauthorrefmark{3}
Australian Centre for Astrobiology, UNSW Australia, Sydney, NSW 2052, Australia}}

\maketitle

\begin{abstract}
The field of exoplanetary science is one of the most rapidly growing areas of astrophysical research. As more planets are discovered around other stars, new techniques have been developed that have allowed astronomers to begin to characterise them. Two of the most important factors in understanding the evolution of these planets, and potentially determining whether they are habitable, are the behaviour of the winds of the host star and the way in which they interact with the planet. The purpose of this project is to reconstruct the magnetic fields of planet hosting stars from spectropolarimetric observations, and to use these magnetic field maps to inform simulations of the stellar winds in those systems using the Block Adaptive Tree Solar-wind Roe Upwind Scheme (\bruu) code. The \bru code was originally written to investigate the behaviour of the Solar wind, and so has been altered to be used in the context of other stellar systems. These simulations will give information about the velocity, pressure and density of the wind outward from the host star. They will also allow us to determine what influence the winds will have on the space weather environment of the planet. This paper presents the preliminary results of these simulations for the star $\tau$ Bo{\"o}tis, using a newly reconstructed magnetic field map based on previously published observations. These simulations show interesting structures in the wind velocity around the star, consistent with the complex topology of its magnetic field. 

\end{abstract}


\section*{Introduction}

Exoplanetary science is one of the most dynamic and rapidly expanding fields in modern astrophysical research. Since the first discovery of a planet orbiting a sun-like star, less than 2 decades ago \cite{Mayor1995}, astronomers have found over fifteen-hundred planets orbiting other stars\footnote{As of 6th November, 2014, the current tally stands at 1516 confirmed planets, with a further 3359 candidate planets awaiting confirmation, according to the Exoplanet Orbit Database at \url{http://exoplanets.org} \cite{Wright2011}.}. The early discoveries were of planets of comparable mass to, or more massive than, Jupiter, and orbiting at distances much closer to their star than the Earth is from the Sun \cite{Marcy1996, Butler1997, Butler1998}. Advances in technology, and the increasing length in the temporal baseline of observations, have resulted in the discovery of systems that have large planets orbiting at similar distances to the gas giants of our Solar system \cite{Howard2010,Wittenmyer2012,Robertson2012}. Also, as a result of the exquisite precision with which the transits of planets across their host stars can now be measured, the size of the smallest discoverable planets has been decreasing, such that the first ``super-Earths'' have now been found \cite{Howard2012, Delfosse2013, Wittenmyer2014}. 

As a result of the dramatic progress made in the discovery of exoplanets, the focus of exoplanetary science has shifted to include trying to characterise the planets that have been found. It will not be long before the first Earth-analogue planets will be discovered, and the hunt for life on those planets will begin. The characterisation of these Earth-like planets will be pivotal in deciding which ones to examine more closely. Determining the habitability of a planet based on a large range of variables is vital to choosing the most promising target for the search for life outside our Solar system \cite{Horner2010}. 

A vital component in determining the habitability of a planet is the influence of the planet's host star. In particular, the activity and winds resulting from the magnetic field of an exoplanet's host-star will play a critical role in ensuring that the planet will be able to sustain a sufficient atmosphere to ensure that life can develop and thrive. In our own Solar system, Mars stands as an example of the destructive power winds can have on a planet's atmosphere. Over the four and a half billion year history of our Solar system, Mars's once thick atmosphere has been stripped by the wind of our Sun \cite{Dehant2007}. Earth has been able to maintain the atmosphere required to support life due to its protective magnetic field. 

It is clear, then, that an understanding of the interaction between the stellar winds of planet host stars and the planets that orbit them will play an important role in the search for life outside our Solar system. However, this area of research is still in its infancy, due to the difficulties involved in either detecting and measuring stellar winds directly or in modelling them based on the information we can obtain about their magnetic field behaviour. 

The first detections of Solar-like winds around other stars were made by examining the blue-shifted absorption of Lyman-$\alpha$ photons by heated atomic hydrogen gas in the Interstellar Medium (ISM) \cite{Wood2004}. This is an indirect method of detecting winds, and attempts have been made to more directly detect winds through looking for x-ray emission from the interaction of the winds and the ISM \cite{Wargelin2002}, or radio emission direct from these winds \cite{Mullan1992}. Unfortunately, these methods have proven to be two to three orders of magnitude less sensitive than the Ly$\alpha$ absorption technique \cite{Wood2004}. As a result of the lack of direct observation, the behaviour of stellar winds remains poorly understood. We are thus reliant on the modelling of these stars, constrained by the observations we can make, to investigate the behaviour of stellar winds.  

In order to model the winds of other stars, we require information on the behaviour of their magnetic fields, and this can be obtained using Zeeman Doppler Imaging (ZDI) \cite{Semel1989}. Making spectropolarimetric\footnote{Spectropolarimetry is an observational technique that involves passing the light from a star through a polariser before sending it into a high-resolution spectrograph. Since magnetic fields will polarise light from atomic spectra, observing polarised spectra then detects the magnetic field of a star. A detailed explanation of this can be found in \cite{Carter1996}.} observations at different phases of a star's rotation, ZDI can reconstruct the topology of the radial, azimuthal and meridional parts of its magnetic field (e.g. \cite{Donati1997c,Donati1999}). 

The purpose of this project is to take the reconstructed radial magnetic fields of planet hosting stars from spectropolarimetric observations, and use these magnetic field maps to inform a simulation of their winds using the \bru code \cite{Powell1999}. In this paper, we will outline our chosen data for this project and target star for initial testing, and briefly discuss the construction of the magnetic field maps that will be used as input for the stellar wind modelling. We will then present some of the preliminary results of the project, and outline the work to be completed in the future. 

\section*{Data Sample}
\label{sec:DataSample}
This project will make use of spectropolarimetric data available through our membership of the BCool\footnote{The BCool project (\url{http://bcool.ast.obs-mip.fr/}) is an international collaboration that is carrying out ongoing observations to categorise and understand the magnetic fields of cool stars. } project \cite{Marsden2014}, with a focus on investigating solar type, planet hosting stars. The first star to be examined for preliminary testing is $\tau$ Bo{\"o}tis (\tbb), a solar type star with a $\sim 5$ Jupiter-mass planet that orbits at a distance of 0.05 AU and with a period of 3.31 days \cite{Butler1997}. Table \ref{tab:tauboo} gives a summary of the basic properties of this star taken from the Exoplanet Orbit Database\footnote{\url{http://exoplanets.org/}. Data were obtained on the 17th October, 2014.} \cite{Wright2011}. 

\tb is an active star, whose magnetic field and wind behaviour has been previously studied (e.g. \cite{Catala2007,Donati2008a,Fares2009,Vidotto2012}), and so is an ideal test target. It displays a cycle in its magnetic field similar to that of our Sun's $\sim 22$ year magnetic cycle, but over a much shorter timescale of around 2 years \cite{Fares2009}. Given the number of magnetic observations taken over many epochs, this makes it an ideal candidate for investigating the temporal evolution of the wind during its magnetic cycle, and the impact this may have on its planet.

\begin{table}[H]
\renewcommand{\arraystretch}{1.3}
\caption{Summary of the basic properties of the star \tbb. }
\label{tab:tauboo}
\centering
\begin{tabular}{|l|l|}
\hline
Mass (Solar masses) & 1.34 \cite{Takeda2007}\\ 
Radius (Solar radii) & 1.46 \cite{Torres2010}\\
Stellar $v \sin i$ (km s$^{-1}$) & 14.98 \cite{Valenti2005}\\
Stellar Inclination Angle (deg.) & 40 \cite{Catala2007}\\
Apparent Magnitude & 4.5 \cite{vanBelle2009}\\
Effective Temperature (K) & 6387 \cite{Valenti2005}\\
Spectral Type & F6IV \cite{Mason2001}\\
\hline
\end{tabular}
\end{table}
\section*{Stellar Magnetic Field Maps}
\label{sec:MagMaps}
The spectropolarimetric observations obtained of our target stars are used to map stellar magnetic fields using ZDI. The process for reducing the data is based on the latest version of the ``Echelle Spectra Reduction: an Interactive Tool'' software \cite{Donati1997b}, called {\sc libre-esprit}, developed by Jean-Francois Donati of the Observatoire Midi-Pyrenees. This software incorporates the optimal extraction of spectra from the CCD images, and uses least squares deconvolution (LSD) to co-add spectral lines, significantly increasing the sensitivity of magnetic detections in the spectra. This is needed, as the magnetic field signal is only $\sim 0.1\%$ of the total light from a star \cite{Donati1997b}. Since the magnetic field will effect the all spectral lines in a similar way, the thousands of spectral lines in each Echelle spectrum can be co-added in this way to improve the signal-to-noise of the magnetic field detection. 

The method of ZDI then uses these LSD profiles from observations made at different phases in the star's rotation to reconstruct the map of the magnetic field for each of the radial, azimuthal and meridional field components\footnote{More details on this can be found in \cite{Donati1997c}.}. Examples of such maps for young, solar type stars can be found in \cite{Marsden2006, Waite2011, Marsden2011}. The stellar winds have been shown to be mainly governed by the radial component of the magnetic field \cite{Jardine2013}, and so the radial field map alone is used as input to the simulations in our work. 

\begin{figure}[H]
\centering
\includegraphics[width=0.9\textwidth]{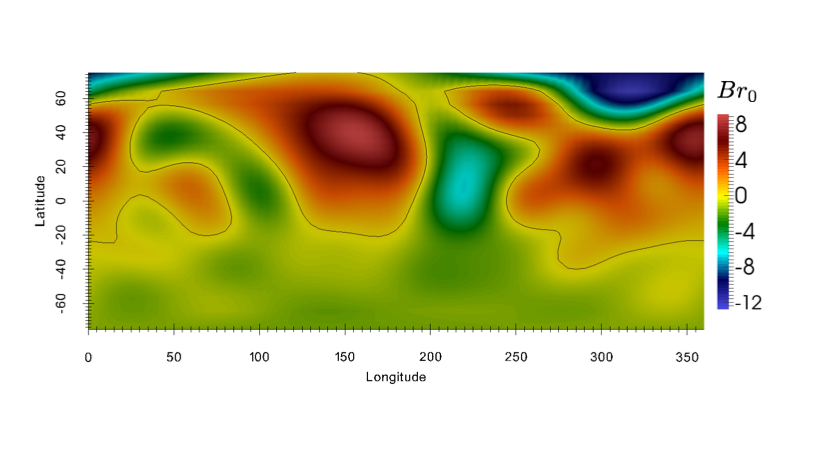}
\caption{A plot of the surface magnetic field of \tb from observation taken in July 2008 by Fares \textit{et al.} \cite{Fares2009}. The black line indicates where the magnetic field is zero. The most interesting feature of this plot are the multiple regions of positive and negative magnetic field, with a large negative spot around the northern pole, and the multiple, large positive regions in the mid-Northern latitudes. No information is available for the far southern region of the star as that is pointed away from the Earth. }
\label{fig:Br0}
\end{figure}

Figure \ref{fig:Br0} shows the radial magnetic field configuration at the surface of \tb from observations made by Rim Fares and collaborators \cite{Fares2009} in July 2008 using the Naval echelle spectropolarimeter on the Telescope Bernard Lyot (TBL) in France. This magnetic field map differs slightly from the those published of the same data set previously due to advances in the ZDI method. The map shows multiple regions of positive and negative magnetic field, with several large positive field spots and a large negative spot around the Northern pole. The field is less detailed at the Southern pole due to a lack of information resulting from the inclination angle of the star with respect to us. The black line indicates where the field is zero. 

\section*{Stellar Wind Modelling With \bruu}
\label{sec:bru}
The wind modelling for our target stars is done using the \bru code, originally developed to simulate the Solar wind \cite{Powell1999}. \bru solves the three-dimensional, ideal magnetohydrodynamic equations:

\begin{equation}
\frac{\partial \rho}{\partial t} + \nabla \cdot (\rho {\mathbf u}) = 0,
\end{equation}
\begin{equation}
\frac{\partial (\rho {\mathbf u})}{\partial t} + \nabla \cdot \left[ \rho {\mathbf u} \otimes {\mathbf u} + \left(P+\frac{B^2}{8\pi}\right)I - \frac{{\mathbf B} \otimes {\mathbf B}}{4\pi} \right]= \rho g,
\end{equation}
\begin{equation}
\frac{\partial {\mathbf B}}{\partial t} + \nabla \cdot \left({\mathbf u} \otimes {\mathbf B} - {\mathbf B}\right) = 0,
\end{equation}
\begin{equation}
\frac{\partial \epsilon}{\partial t} + \nabla \cdot \left[ {\mathbf u} \left( \epsilon + P + \frac{B^2}{8\pi} \right) - \frac{({\mathbf u} \cdot {\mathbf B}){\mathbf B}}{4\pi}\right] = \rho {\mathbf g} \cdot {\mathbf u},
\end{equation}

which describe the conservation of mass, momentum, magnetic flux and energy, respectively. In these equations, $\rho$ is the mass density, ${\mathbf u}$ is the plasma velocity, ${\mathbf B}$ is the magnetic field, $P$ is the gas pressure, ${\mathbf g}$ is the gravitational acceleration of a star with mass $M_*$ and radius $R_*$, $I$ is the identity matrix, and $\epsilon$ is the total energy density, given by

\begin{equation}
\epsilon = \frac{\rho u^2}{2} + \frac{P}{\gamma -1} + \frac{B^2}{8 \pi}.
\end{equation}

Here, $\gamma$ is the polytropic index such that $P \propto \rho^{\gamma}$.

\bru has been chosen for this work as it has been used extensively \cite{Ulusen2010,Vidotto2011,Kozarev2013}, and reliably models the wind behaviour of the Sun. Additionally, the block refinement within the code allows the resolution to be increased in specific areas of interest, minimising the load on computational resources by not having to increase resolution everywhere. It is a modular code, designed to be used in different ways and with different physical parameters than the original solar parameters. In essence the code modifications involve the replacement of physical constants such as solar mass and radius sun by stellar parameters that can be varied as needed for each new stellar target.  For more details see \cite{Vidotto2009,Vidotto2010}.

\section*{Preliminary Results and Discussion}
\label{sec:results}
\begin{figure}[H]
\centering
\includegraphics[width=\textwidth]{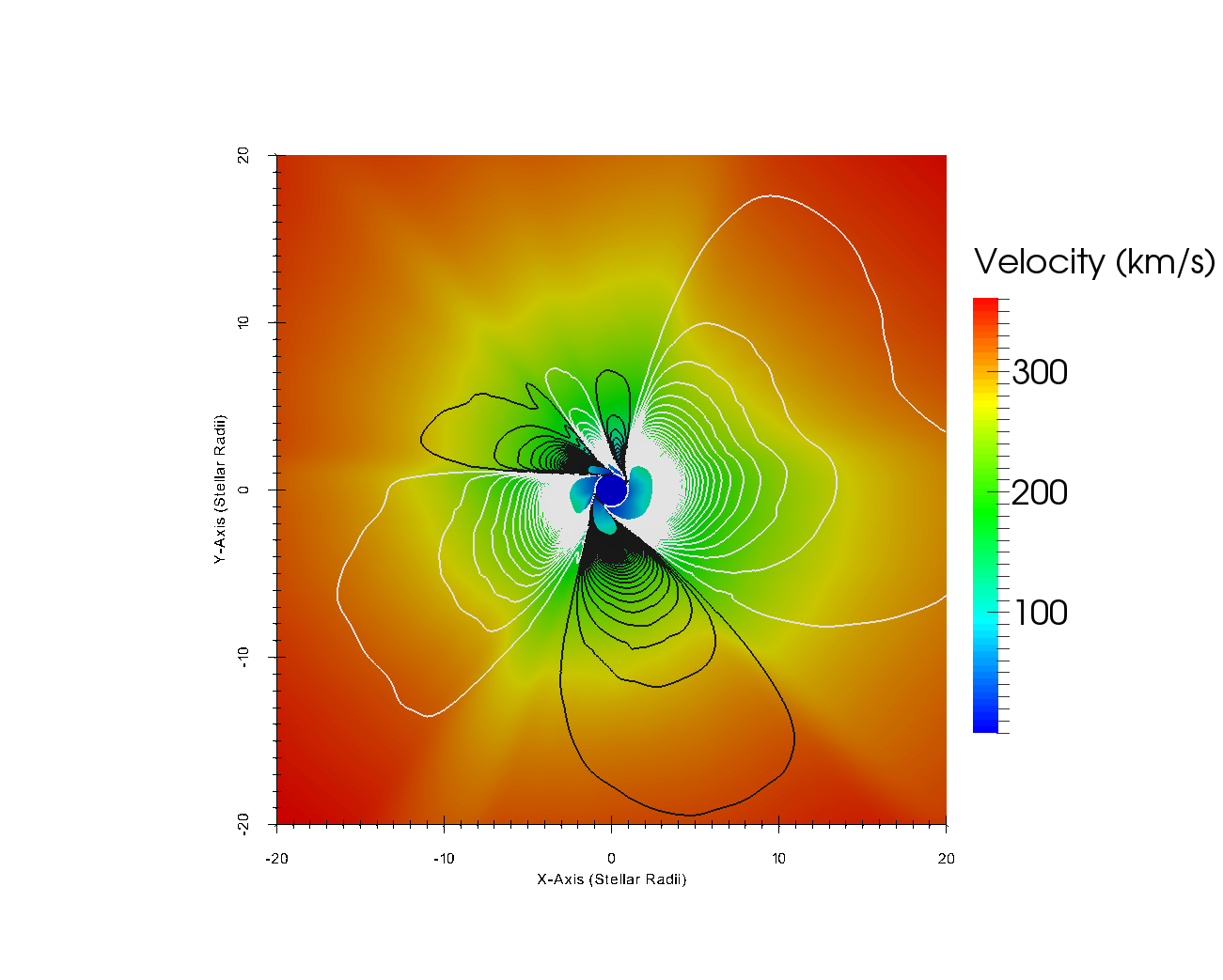}
\caption{A plot of the wind velocity of the star \tb from observations taken in July 2008 by Fares \textit{et al.} \cite{Fares2009}, in the equatorial plane of the star out to 20 stellar radii. Overlaid are contours of radial magnetic field ranging from -0.2 to +0.2 Gauss, with white indicating positive field and black indicating negative field. Showing these contours highlights the interesting `streams' of higher wind velocity at the boundaries between positive and negative magnetic field.}
\label{fig:Umag}
\end{figure}

Some preliminary results from our implementation of the \bru code for the star \tb are detailed in Figure \ref{fig:Umag}. The plot shows a slice in the equatorial plane of the star out to 20 stellar radii, with the colour map indicating the magnitude of the wind velocity in km s$^{-1}$. For context, contours of radial magnetic field from -0.2 to +0.2 Gauss are overlaid in black (negative field) and white (positive field). The complex magnetic field structure seen in the lobes of positive and negative field contours produces abrupt differences in wind velocities around the star. Further work is needed to confirm that the apparent sharp changes in wind velocity are completely physical, and not an artefact of the limited phase coverage of the magnetic mapping, or the limited resolution of the wind modelling used to produce these preliminary results. However, these `streams' of higher velocity wind at the boundary between the areas of positive and negative magnetic field are particularly interesting, and could suggest the funnelling of plasma through these boundary regions. 

A future goal is to understand the temporal behaviour of the star \tbb, as well as understanding the impact its planets will have on the behaviour of the wind. We will explore further targets, including others that have been mapped with ZDI, which will be used to investigate the time evolution of winds,  and some that have not previously been explored with ZDI. 

As well as producing wind models for the target stars, we will modify the \bru code to simulate the interaction of the stellar wind with a planet for each target. Beyond allowing us to better characterise these particular planetary systems, these simulations will also help us to better understand the formation and evolution of planetary atmospheres beyond our Solar system. 

It is through that work that we aim to understand the impact of the winds on the planets that orbit these stars, and similarly how the presence of a planet, which may or may not have a magnetic field of its own, impacts on the behaviour of the wind.

\section*{Acknowledgments}
This research is supported by the University if Southern Queensland's Strategic Research Fund StarWinds project. This research has made use of NASA's Astrophysics Data System. This research has made use of the Exoplanet Orbit Database and the Exoplanet Data Explorer at \url{http://exoplanets.org}.



\bibliographystyle{IEEEtran}
\bibliography{ASRC14_bib}
%
%
%
%
%
%
%
%

\end{document}